\documentclass[multphys,vecphys]{svmult}

\usepackage{makeidx}         
\usepackage{graphicx}        
\usepackage{multicol}        
\usepackage[bottom]{footmisc}

\makeindex             

\begin{document}

\title*{The radii of thousands of star clusters in M51 with \emph{HST/ACS}}
\author{R. A. Scheepmaker\inst{1}\and
 M. Gieles\inst{1}\and M. R. Haas\inst{1}\and N. Bastian\inst{2}\and S. S. Larsen\inst{1}\and H. J. G. L. M. Lamers\inst{1,3}}
\authorrunning{R. A. Scheepmaker\inst{1}\and M. Gieles\inst{1}\and M. R. Haas\inst{1} et al.}
\institute{Utrecht University
\texttt{scheepmaker@astro.uu.nl}
\and University College London \and SRON Laboratory for Space Research}
%
\maketitle

\section{M51 -- a star cluster laboratory}
\label{sec:M51 -- a star cluster laboratory}

The young ($<1$~Gyr) star cluster population of M51 (NGC 5194, Hubble-type Sc) is, due to its relative proximity and high number of star clusters, an excellent candidate for studies to the formation and evolution of young star clusters (YCs) in spiral disks.
M51 has many star clusters because of its high star formation rate, which may have been caused by its interaction with NGC 5195, and because the disk is almost face-on, we can observe the clusters throughout the disk without being hampered by extinction too much.
The new \emph{HST/ACS} mosaic image of M51, taken as part of the Hubble Heritage program \cite{mutchler05}, covers the entire disk in 6 pointings in \emph{F435W} ($\sim B$), \emph{F555W} ($\sim V$), \emph{F814W} ($\sim I$) and \emph{F658N} ($H\alpha$), with a resolution of $\sim$2~pc per pixel.
Previous studies on the star cluster population of M51 used lower resolution \emph{WFPC2} data that did not cover the entire disk (see e.g.~\cite{larsen00,bastian05a,lee05}).
In the study of \cite{bastian05a}, clusters were selected based on their fit to SSP models, which led to a possible contamination of individual stars.
In a different study clusters were selected based on their sizes, which made it difficult to select a large sample due to the limited resolution \cite{lee05}. 
However, by exploiting the superb resolution of the \emph{ACS} camera and the large field of view of $\sim 17.5\times24.8$~kpc, we now \emph{can} select a large sample of clusters based on their sizes.
This sample can then be used to study the formation history and evolution of the star cluster population and to derive infant mortality rates and study how these depend on the environment.
Since this is work in progress, we here select a smaller preliminary sample of clusters and we present some first results that can already be derived by looking at the sizes and the positions of the star clusters in the disk of M51.

\section{Selecting a reliable cluster sample based on radii}
\label{sec:Selecting a reliable cluster sample based on radii}

We measure the effective radii of $75\,436$ sources in our data with the \emph{Ishape} routine \cite{larsen99}, which convolves the PSF of the telescope with Moffat 15 profiles \cite{moffat69} of different sizes and then selects the fit with the lowest $\chi^2$.
In this way \emph{Ishape} can determine an accurate radius of a marginally extended source down to $\sim$0.5~pc at the distance of M51.
We assume that a source is a cluster when its radius is $>0.5$~pc and when a fit using a Moffat profile gives a lower $\chi^2$ than a fit using just the PSF (i.e. assuming the source is a point source).
This gives us a sample of $14\,950$ clusters.
If we apply a 90\% completeness limit of $B$ \& $V$ $<$~23.3 mag, which we determined for a 3~pc source in a high background region, we find a sample of 4357 resolved clusters above the completeness limit.
This sample was used to study the cluster luminosity function and the related maximum mass for star clusters in M51 (\cite{gieles06}, also Gieles et al.\ in these proceedings).

This sample can also be used to study the distribution of the clusters in the disk of M51 (\S\ref{sec:The distribution of star clusters in the disk}).
However, it turns out that many of the clusters in this sample are in crowded regions or in regions with a highly varying background, making the value of the measured radius unreliable due to blending effects and fits to parts of the background light.
We therefore also select a smaller sample of 769 clusters which are isolated and in low background regions and therefore have more accurate size estimates.
We use this accurate cluster sample to study the radii themselves (\S\ref{sec:A preferred radius for both young and globular clusters} and \S\ref{sec:Radius versus distance}).

\section{The distribution of star clusters in the disk}
\label{sec:The distribution of star clusters in the disk}

We use the sample of 4357 resolved clusters to plot the surface density distribution of the clusters in the disk of M51.
Fig.~1.a. shows a bump in the surface density distribution at a galactocentric distance of $\sim$6~kpc.
This distance is remarkably similar to the corotation radius of $\sim$5.5~$\pm$~1.0~kpc, which is the distance where the rotational velocity of the stellar and gaseous component is the same as the one of the spiral arms \cite{zimmer04,tully74}.
The bump at this location indicates that the corotation radius is a preferred site for cluster formation.
We note that one of the biggest complexes of clusters, named G2, is also found near the corotation radius \cite{bastian05b}. 

Assuming the surface density distribution of clusters is an exponential distribution of the form $\Sigma \propto e^{-D/R_{h}}$, we can derive the scalelength $R_{h}$.
Stars in the disk of the Milky Way (MW) follow an exponential distribution with a scalelength of $\sim$3.5~kpc \cite{BT87}.
For M51 this scalelength is expected to be larger, since the disk of M51 is bigger than the disk of the MW.
Due to the dissipative nature of molecular gas, the distribution of GMCs is in general more centrally concentrated than the distribution of the stars.
The surface density of GMCs in M51 has a scalelength of $R_{h} = 2.4$~kpc \cite{garcia93}. 
Since we are looking at \emph{young} clusters, we would expect them to be correlated with the GMCs more than the total field population.
In Fig.~1.a. we overplotted the surface density distributions of GMCs and stars.
Obviously, because of the presence of the bump, the surface density distribution of the clusters can not be approximated by a straight line.
However, the clusters seem to be more correlated with the GMCs, especially for distances $<$3~kpc.
In a future paper we will address this issue in much more detail, also by measuring the surface distribution of the stars in M51 ourselves \cite{scheepmaker06}.
   
\begin{figure}
\centering
\includegraphics[height=4cm]{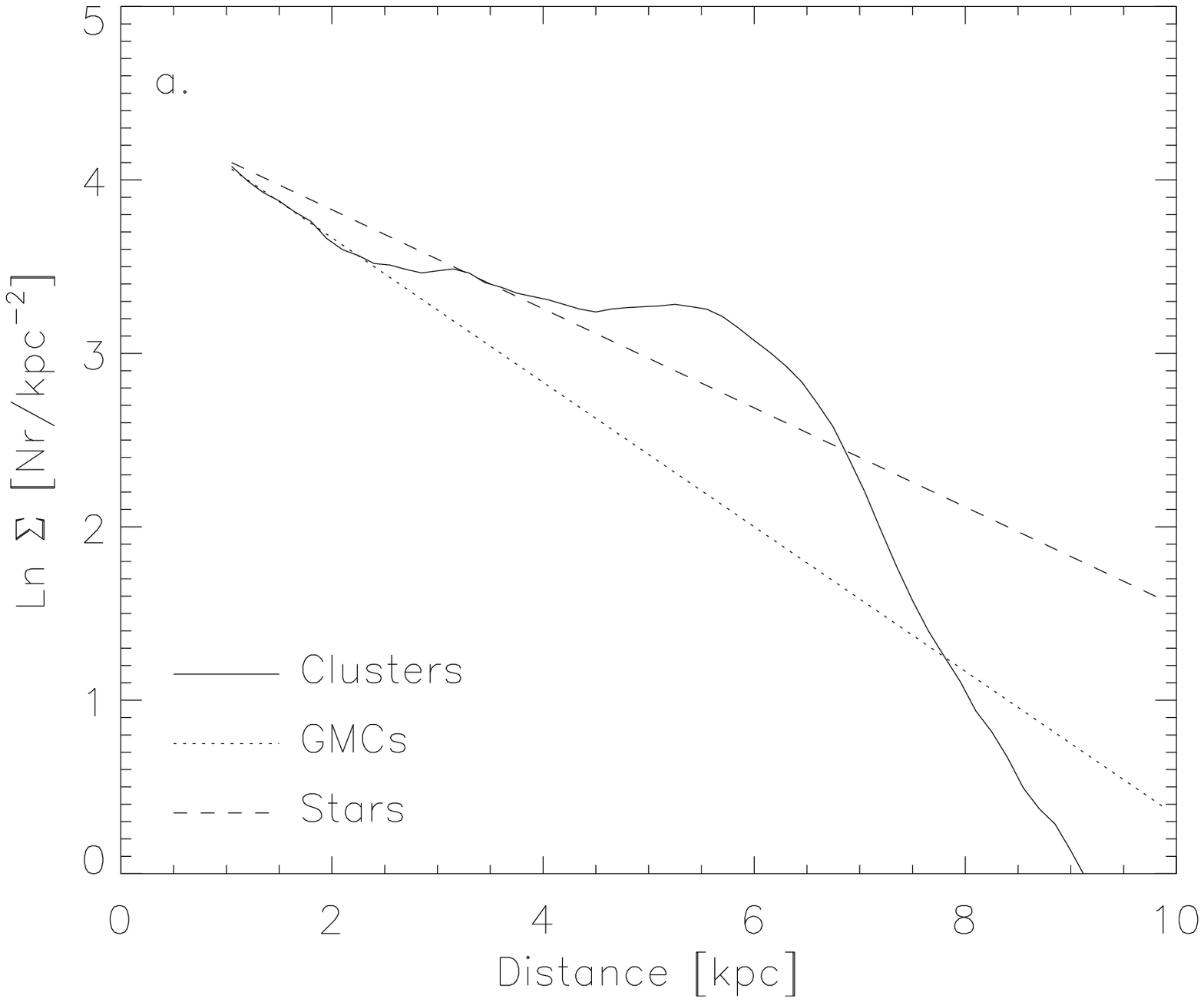}
\includegraphics[height=4cm]{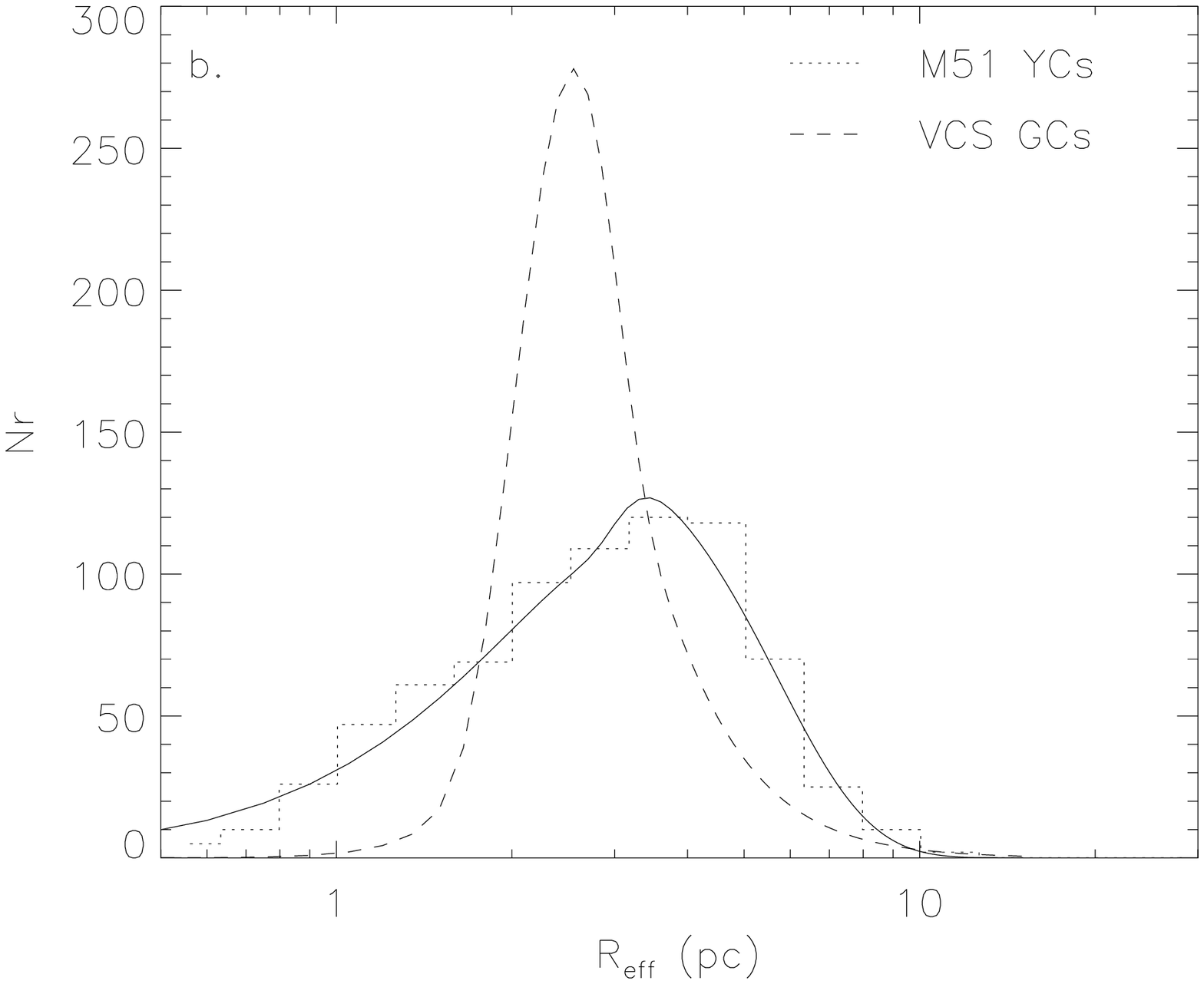}
\caption{{\bf a.} The surface density versus galactocentric distance for the complete sample of 4357 resolved clusters in M51 (solid line). The dashed line shows the density distribution of stars in the Milky Way ($R_{h} = 3.5$~kpc) and the dotted line shows the density distribution of GMCs in M51 ($R_{h} = 2.4$~kpc). {\bf b.} The effective radius distribution of 769 star clusters in M51 (histogram). The solid line is the best fit of Eq.~22 of \cite{jordan05}. The dashed line is the best fit of the same function to the GCs of the \emph{ACS} Virgo Cluster Survey.}
\label{fig:surface density distribution}       
\end{figure}

\section{A preferred radius for both young and globular clusters}
\label{sec:A preferred radius for both young and globular clusters}

In Fig.~1.b. we show the effective radius distribution of our smaller sample of 769 clusters, for which we have determined more accurate radii (\S\ref{sec:Selecting a reliable cluster sample based on radii}). 
Also plotted is the radius distribution of the much older GCs of the \emph{ACS} Virgo Cluster Survey (\emph{VCS}) of \cite{jordan05}.
Both distributions, concerning completely different populations, have their peak at $\sim$3~pc, while the general shapes of the distributions are different.
For early-type galaxies, the mean effective radius scales inversely with galaxy color \cite{jordan05}.
It is not clear if the same relation also holds for late-type spiral galaxies, but when we correct the radii of the \emph{ACS VCS} GCs to the color of M51, the peak shifts to the right and the similarity in the location of both peaks is even stronger.
The effective radius distribution of the galactic GCs also peaks at 3~pc \cite{harris96}.
This suggests that for both young and old populations of star clusters there exists a preferred effective radius of $\sim$3~pc.
The preferred radius also implies that the procedure for deriving distances, based on the median radius and described in \cite{jordan05}, could also work for spiral galaxies with young cluster populations. 
If we apply the procedure to our cluster sample we find a distance of $8.6\pm0.9$~Mpc, very close to the assumed distance to M51 of $8.4\pm0.6$~Mpc based on the planetary nebulae luminosity function \cite{feldmeier97}.

\section{Radius versus distance}
\label{sec:Radius versus distance}

The GCs in the Milky Way show a relation between their effective radius and galactocentric distance of the form $R_{\mathrm{eff}} \propto D^{\sim1/2}$ \cite{vandenbergh91}, which can be explained by both a remnant of the formation process (denser clouds near the Galactic centre) and by tidal relaxation of the clusters in the Galactic potential, since the tidal radius scales as $\propto D^{2/3}$.
Our M51 clusters show \emph{no} relation between radius and galactocentric distance, meaning that they did not form in tidal equilibrium.
It also means that the radius of the clusters is not related to the ambient pressure in their host galaxy, since this pressure decreases with galactocentric distance.

\section{Conclusions}
\label{sec:Conclusions}

We measure the radii of $75\,436$ sources in M51 to select and study a large cluster sample covering the complete spiral disk.
Some first preliminary results show a hint of enhanced cluster formation at the corotation radius and a preferred radius of $\sim$3~pc for YCs, which is similar to the preferred radius of much older GCs.
However, in contrast to the GCs, the YCs in M51 do not show a relation between radius and galactocentric distance. 
\\ \\
\emph{Acknowledgement.} RS gratefully thanks Andr\'es Jord\'an for help in fitting our radius distribution with his function.



\printindex
\end{document}